\documentstyle[aps,prb,twocolumn,psfig,floats]{revtex}
\begin{document}
\draft
\title{Mean-Field Theory for the Spin-Triplet Exciton Liquid
in Quantum Wells}
\author{R. J. Radtke and S. Das Sarma}
\address{Center for Superconductivity Research, Department of Physics,\\
University of Maryland, College Park, Maryland  20742-4111}

\address{\mbox{ }}
\address{\mbox{ }}
\address{\parbox{16cm}{\rm \mbox{ }\mbox{ }\mbox{ }
Using a mean-field theory, we study the possible existence of
a spin-triplet intersubband exciton liquid ground state
in semiconductor quantum well systems
as a function of the electronic density and the strength of the
intersubband Coulomb interaction matrix element at low temperatures.
We find the excitonic phase to be stable over a large region of
parameter space, and our calculated critical
temperatures are attainable experimentally.
In addition, we find that the transition to the excitonic phase can
be either first- or second-order at zero temperature.}}

\maketitle

\makeatletter
\global\@specialpagefalse
\def\@oddhead{REV\TeX{} 3.0\hfill Das Sarma Group Preprint, 1995}
\let\@evenhead\@oddhead
\makeatother

\narrowtext

Interest in the effects of electronic correlations in low-dimensional
quantum systems such as quantum wells and quantum wires is
motivated by both applied and basic science objectives.
On the applications side, the continuing reduction in the
characteristic size of the components of integrated circuits
suggest that quantum effects will eventually become important
design considerations in these devices.
In addition, the unusual electronic properties of low-dimensional
quantum systems present the possibility of developing new
devices to exploit these properties.
On the basic science side, low-dimensional quantum systems
are almost ideal realizations of simple quantum mechanical
systems, and so are test-beds for our understanding of
single- and many-particle aspects of quantum mechanics.
In particular, the ability to enter a regime where
electronic correlation effects become large opens the possibility
of discovering new, strongly correlated quantum
phases, as has already happened
with the fractional quantum Hall effect.\cite{fqhe}
In this paper, we describe another possible correlation-driven
phase which occurs in low-dimensional quantum systems in the
absence of a magnetic field:  the spin-triplet exciton liquid.

The existence of this phase was recently predicted theoretically by Das
Sarma and Tamborenea based on a local-density approximation (LDA)
calculation of the collective mode spectra of either double or
wide single quantum wells.\cite{Pablo}
In these calculations, the frequency of the long-wavelength intersubband
spin-density excitations vanishes for electron densities around
10$^{10}$ to 10$^{11}$ cm$^{-2}$.
This mode softening implies that intersubband
spin-density excitations are spontaneously generated in the system
and thus signals the presence of a new phase.\cite{Pablo}
Since these excitations involve, for example,
a spin-up electron in one subband and a spin-down hole in
another, this new phase has been termed a spin-triplet exciton liquid.

\begin{figure}[t]
\vskip 3\baselineskip
\psfig{file=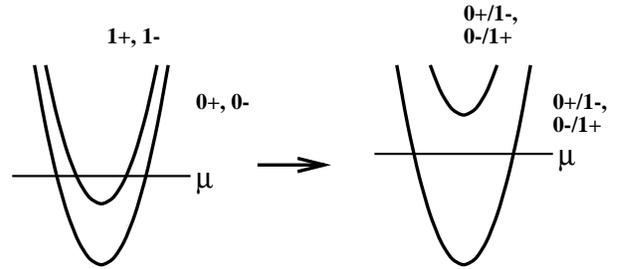,width=3.125in,angle=90}
\caption{Change in the electronic dispersion from the normal phase
with two subbands occupied (left side of the figure) to the excitonic
phase (right side of the figure).
In the normal phase, the band structure consists of two subbands
(0 and 1), each with degenerate spin-up (+) and spin-down (-) components;
in the excitonic phase, the band structure still contains two
subbands, but they are linear
combinations of the original bands that result in a larger
subband splitting and the occupation of only the lowest subband.
In both cases, all states below the chemical potential $\mu$ are filled
at zero temperature.}
\label{fig:spectrum}
\end{figure}

In bulk semiconductors, the transition to an exciton liquid has
been studied extensively.
Previous theoretical work examined the formation of an
excitonic insulator near a semimetal-semiconductor transition,
\cite{Mott,Knox,KK,Cloizeaux,KM,Rice}
an excitonic metal due to Fermi-Dirac condensation,
\cite{FermiDirac} and
an excitonic superfluid resulting from a Bose-Einstein condensation.
\cite{Blatt,Keldysh}
However, bulk semiconductors possess a three-dimensional structure
and a (usually indirect) energy gap between
the conduction and valence bands.
By contrast, we consider here the conduction subbands
in modulation-doped two-dimensional semiconductor microstructures
which do not have
a gap in their (two-dimensional) single-particle excitation spectra
[cf. the left side of Fig.~\ref{fig:spectrum}].
Thus, in order to describe the phase predicted by Das Sarma and
Tamborenea, it is necessary to reconsider the theory of the exciton
liquid within the context of quantum wells.\cite{note}

As yet, no definitive experimental evidence for the existence
of the excitonic phase has been presented; indeed, beyond the
softening of the intersubband spin-density excitations, no
calculations exist about the properties of the new phase.
However, the collective mode softening should be observable by
inelastic light scattering techniques.\cite{Raman}
Moreover, the use of the LDA to compute the collective modes in
quantum wells in both the charge and spin channels is well
established\cite{modes} and yields good agreement with experiment,
even when many-body effects become important.\cite{Decca}
We therefore expect that the excitonic phase should be observed
experimentally in the future.
In anticipation of this discovery, we present here a brief description
of the spin-triplet exciton phase based on a mean-field theory.
We determine the nature of the new phase and establish its boundaries
in temperature, density, and intersubband Coulomb repulsion.
A more detailed description of these calculations, along with a
computation of the collective modes and the effect of impurities
on the excitonic phase, will appear elsewhere.\cite{radtke}

The physical system we wish to describe consists of an
electron gas in the presence of a confining potential along the
$z$-direction which interacts via a screened Coulomb repulsion.
The confining potential may be viewed as arising from either a
double quantum well or a wide single quantum well; in either case,
the two lowest-energy bound states of the confining potential are
well separated in energy from the next-lowest energy level.
This situation can arise, for example, in the lowest
symmetric-antisymmetric subband splitting in a double quantum
well structure.
Since we wish to focus on the general features of the condensed
phase and not on the detailed effects arising from the wave-vector
and frequency dependence of the screened Coulomb interaction, we
take the interaction to be a short-range repulsion.
In principle, we may include more detailed forms of the interaction
within our mean-field formalism, but these detailed calculations
should not affect the qualitative features we will discuss.
Furthermore, the short-range repulsion model has the
advantage of being a single-parameter description of the interaction
and should
be a reasonable approximation at densities which are not too low.

We work in the basis defined by the product of eigenstates of
the confining potential $\xi_\alpha (z)$ and two-dimensional
free particles $e^{i{\bf k \cdot r}} / \sqrt{V_0}$.
In this basis, the Hamiltonian may be written
\begin{eqnarray}
H &=& \sum_{\alpha {\bf k} \sigma} \,
  \left[ E_\alpha - \frac{\hbar^2 {\bf k}^2}{2m} - \mu \right]
  c^{\dag}_{\alpha {\bf k} \sigma} c^{ }_{\alpha {\bf k} \sigma}
  \nonumber \\
&+& \frac{1}{2 V_0} \, \sum_{\alpha_i {\bf k}_i {\bf q} \sigma_i}
  \, V_{\alpha_1 \alpha_4, \alpha_2 \alpha_3} \nonumber \\
  && \times c^{\dag}_{\alpha_1 {\bf k_1 + q} \sigma_1}
  c^{\dag}_{\alpha_2 {\bf k_2 - q} \sigma_2}
  c^{ }_{\alpha_3 {\bf k_2} \sigma_2}
  c^{ }_{\alpha_4 {\bf k_1} \sigma_1} ,
\label{eq:h}
\end{eqnarray}
where
\begin{eqnarray}
V_{\alpha_1 \alpha_4, \alpha_2 \alpha_3} =
  V \int dz \, \xi^{*}_{\alpha_1} (z) \xi^{*}_{\alpha_2} (z)
  \xi^{ }_{\alpha_3} (z) \xi^{ }_{\alpha_4} (z)
\end{eqnarray}
is the matrix element of the short-range interaction $V$,
$V_0$ is the area (2D volume) of the confined electron gas,
$\mu$ is the chemical potential,
$E_\alpha$ is the subband eigenenergy of the bound state $\alpha$ of the
confining potential,
and $c_{\alpha {\bf k} \sigma}$ annihilates an electron of effective
mass $m$ in subband $\alpha$ with wave vector ${\bf k}$ and spin
projection $\sigma$.

To make the calculations more tractable,
we take advantage of several
symmetries in the problem and make some simplifying approximations.
First, the two lowest subbands are by assumption well separated
from the others, so we consider only these subbands,
as depicted by the left diagram in Fig.~\ref{fig:spectrum}.
This should be an excellent approximation for energies and
temperatures much less than the splitting between the lowest
and third lowest subbands.
Second, we may take the wave functions $\xi_\alpha (z)$ to be real,
so that the ordering of the indices in the Coulomb matrix
elements is irrelevant.
Third, we make the reasonable assumption that the confining potential
is even in $z$, so that $V_{01,11}$ = $V_{10,00}$ = 0
($\alpha$ = 0 (1) denotes the (next to) lowest subband).
The remaining independent matrix elements are then
$V_{00,00}$, $V_{11,11}$, and $V_{00,11} \equiv V_{01}$.
Finally, for the problem in which we are interested, we may
set $V_{00,00} = V_{11,11} = 0$ without loss of generality.
These diagonal matrix elements act to renormalize the subband splittings
and give rise to ferromagnetic phases at low densities.
While our formalism can easily incorporate these matrix
elements, we are primarily interested in the excitonic phase, so
we leave the detailed effect of these matrix elements
as a subject for future work.\cite{radtke}

\begin{figure}[t]
\vskip 2\baselineskip
\psfig{file=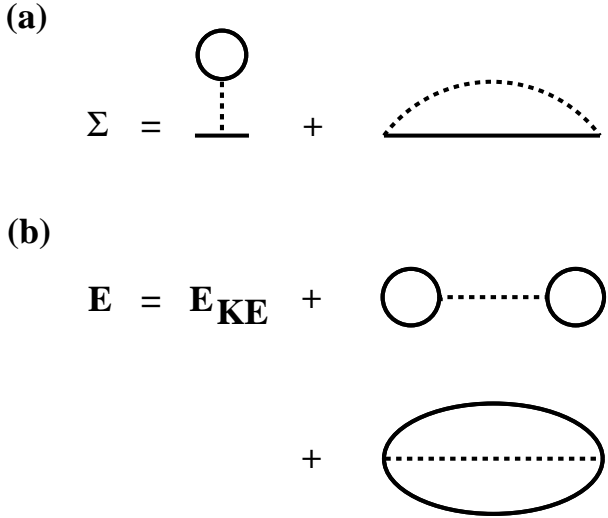,width=3.125in,angle=90}
\caption{Many-body diagrams for (a) the self-energy $\Sigma$ and
(b) the total energy $E$ used in the mean-field theory discussed
in the text.
The solid lines correspond to dressed electronic Green's functions
and the dashed lines to the screened Coulomb interaction;
$\rm E_{KE}$ is the kinetic energy of the interacting electronic system.
In solving the mean-field theory, we approximate the screened Coulomb
interaction by an on-site, repulsive interaction.
See text for details.}
\label{fig:diagrams}
\end{figure}

We solve this two-subband model within mean-field theory.
The propagators and self-energies are defined in the usual way
as matrices in the subband index and spin projection.
With these quantities, we set up and solve the one-loop
self-consistent Hartree-Fock equations depicted by the diagrams of
Fig.~\ref{fig:diagrams}(a).
We simultaneously impose the constraint that the chemical potential
yield a fixed electronic density $n$.\cite{fixedn}
In evaluating the mean-field theory, we allow for the possibility
of self-energies which are off-diagonal in both spin and band
indices.
Since the LDA calculations of Das Sarma and Tamborenea indicate that
the intersubband spin-density excitations soften in the
excitonic phase,\cite{Pablo} we expect that thermal averages
of the form $\Delta_{\alpha \sigma}$ =
$\langle c^{\dag}_{\alpha {\bf k} \sigma}
c^{ }_{1-\alpha, {\bf k}, -\sigma} \rangle$ will
become non-zero in the excitonic phase.
In the language of critical phenomena, $\Delta_{\alpha \sigma}$ is
the order parameter of the phase transition, and
$\Delta_{\alpha \sigma} \neq 0$ indicates the spontaneous breaking
of the intersubband spin symmetry which leads to the excitonic phase.

Indeed, the resulting mean-field equations have two classes of solutions:
a paramagnetic phase where $\Delta_{\alpha \sigma}$ = 0 with
either one or both subbands occupied, and an
excitonic phase where $\Delta_{\alpha \sigma} \neq$ 0.
Direct calculation of the order parameter in the excitonic
phase shows that it is independent of $\alpha$ and $\sigma$ and
may be written as $\Delta$ = $\sqrt{(n V_{01} / 2)^2 - \Delta_{\rm SAS}^2}$,
where $\Delta_{\rm SAS} = E_1 - E_0$ is the splitting between the
lowest two subbands.
Taking the expectation value of the Hamiltonian [Eq.~(\ref{eq:h})]
in the interacting ground state and evaluating it within the
mean-field approximation [Cf. Fig.~\ref{fig:diagrams}(b)],
we find that the excitonic phase is stable when
$2 N_0 V_{01} \geq 1$ and $n V_{01} \geq 2 \Delta_{\rm SAS}$.
(In this expression, $N_0 = m / 2 \pi \hbar^2$ is the
two-dimensional, single-spin density of states).
We also find that, despite the spin-triplet nature of the order parameter,
there is no net magnetic moment in the excitonic phase, similar
to what happens in the Balian-Werthammer description of the B-phase
of superfluid $^3$He.\cite{BW}

Examining the excitation spectrum of the excitonic ground state
shows that it is characterized by a re-arrangement
of the non-interacting bands, as shown in Fig.~\ref{fig:spectrum}.
The interacting bands become linear combinations of, for example, the
spin-up portion of subband 0 and the spin-down portion of subband 1.
These combinations are expected from the form of the order parameter
$\Delta_{\alpha \sigma}$ and may also be obtained from a Bogoliubov
transformation of the original Hamiltonian.\cite{Bogoliubov}
The new combinations increase the effective
splitting between the subbands and so reduce the potential energy
associated with the intersubband Coulomb repulsion.
Moreover, if the second subband is occupied in the normal phase,
the de-population of the second subband in the excitonic phase
reduces the kinetic energy of the system.
These two effects stablize the excitonic phase when the intersubband
Coulomb interaction is sufficiently large.
We note that the interacting single-particle
density of states is not gapped in the excitonic phase,
so the new ground state is neither an insulator nor a BCS-like
superconductor.
Thus, the phase transition to the spin-triplet exciton liquid
may be difficult to observe in transport measurements.

\begin{figure}
\psfig{file=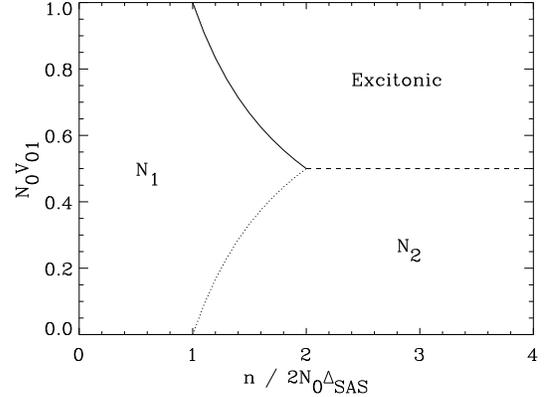,width=3.125in,angle=90}
\caption{Zero-temperature mean-field phase diagram for the
quantum well system described in the text as a function of
the normalized intersubband Coulomb matrix element $V_{01}$
and the normalized electronic density $n$.
The regions correspond to the normal phase with one subband
occupied ($N_1$), the normal phase with both subbands occupied
($N_2$), and the excitonic phase.
Note that the transition from $N_1$ to the excitonic phase is
second-order (solid line), but the transition from $N_2$ is
first-order (dashed line).
In the figure, $N_0 = m / 2\pi \hbar^2$ is the single-spin,
2D density of states, and $\Delta_{\rm SAS}$ = $E_1 - E_0$
is the splitting of the lowest two subbands in the normal phase.}
\label{fig:phase}
\end{figure}

At zero temperature, these results may be summarized in
the phase diagram of Fig.~\ref{fig:phase}.
There are three regions of interest in this figure:
$N_1$, corresponding to the normal phase with only one subband
occupied; $N_2$, corresponding to the normal phase with both
subbands occupied; and the excitonic phase.
As may be seen from the expression for the order parameter,
the phase transition from $N_1$ to the excitonic phase is second
order, but the transition from $N_2$ is {\it first}-order.
Moreover, the excitonic phase always exists for sufficiently large
intersubband repulsion $V_{01}$.

\begin{figure}
\psfig{file=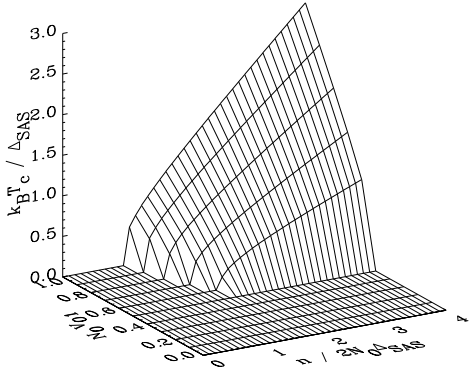,width=3.125in,angle=90}
\caption{Normalized critical temperature $T_c$, below which the
excitonic phase obtains, as a function of the normalized
intersubband Coulomb matrix element $V_{01}$
and the normalized electronic density $n$.
In the figure, $N_0 = m / 2\pi \hbar^2$ is the single-spin, 2D
density of states, and
$\Delta_{\rm SAS}$ = $E_1 - E_0$ is the normal-state splitting
of the lowest two subbands.
Since $\Delta_{\rm SAS} / k_B$ is usually on the order of 10~K or more,
the excitonic phase may be experimentally observable in
suitably constructed quantum wells.}
\label{fig:tc}
\end{figure}

We have also examined the finite-temperature properties of the
mean-field theory and find that the excitonic phase is stable
up to a finite temperature $T_c$ which can be several times
$\Delta_{\rm SAS} / k_B$.
This critical temperature is plotted as a function of the
electronic density $n$ and the intersubband repulsion $V_{01}$
in Fig.~\ref{fig:tc}.
Since $\Delta_{\rm SAS} / k_B$ can be on the order of 10~K or more,
the excitonic phase should be experimentally observable.
We note that our calculation of $T_c$ includes all components of the
self-energy and so is analogous to the strong-coupling theory of
superconductivity;\cite{AM} consequently, our calculated $T_c$ is
the best estimate of the actual critical temperature available
within mean-field theory.
However, since mean-field theories are known to overestimate
critical temperatures in two-dimensional systems,
we cannot rule out the possibility of having very
low transition temperatures in real systems.

We have treated $V_{01}$ and $n$ independently in this
calculation, but in actual systems $V_{01}$ is determined by $n$
and the geometry of the quantum wells.
To make a quantitative comparison of our theory with experiment
therefore requires a detailed calculation of the intersubband
Coulomb matrix element $V_{01}$.\cite{renorm}
This computation may be accomplished with an LDA theory (see, for
example, Ref.~\onlinecite{Pablo}) and will result in a particular
path within the parameter space of the phase diagram of
Fig.~\ref{fig:phase}.
Whether the excitonic phase is observable for that geometry then
becomes a question of detail; specifically, does the resulting
trajectory pass into the excitonic regime?
The calculations of Ref.~\onlinecite{Pablo} suggest that many
quantum well systems can enter the excitonic phase for low
but obtainable electronic densities, so our calculations should
be relevant to these systems.

To summarize, we have developed a mean-field theory which
describes a new phase in two-dimensional quantum systems,
namely the spin-triplet intersubband exciton liquid.
This phase should appear in double or wide single quantum wells
when the intersubband Coulomb interaction is
large, the intersubband energy separation is small,
the electronic density is low, and at temperatures
which may be accessible experimentally.
We have constructed the mean-field phase diagram for the excitonic
phase assuming that the intersubband repulsion and the electronic
density may be varied independently, and we find regions of
both first- and second-order phase transitions at zero temperature.
Further experimental and theoretical work is required in order to
observe and quantify more precisely the properties of this phase
and its relationship to other phases of low-dimensional
systems such as Wigner crystals and ferromagnets.


The authors would like to thank P. I. Tamborenea for stimulating
conversations during the course of this work.
This work was supported by the NSF, the ONR, and the ARO.


\begin{references}
\bibitem{note}One advantage of studying excitonic phases in
artificial microstructures is the tunability of these systems:
these microstructures may be engineered to have a wider range of
physical properties than more traditional three-dimensional materials.
\bibitem{fixedn}This constraint is particularly important in
finite-temperature calculations whenever $k_BT$ is greater than
or of order $\Delta_{SAS}$ = $E_1 - E_0$.
\bibitem{renorm}Note that the interaction matrix element $V_{01}$
and the subband splitting $\Delta_{\rm SAS}$ which enter our
theory are renormalized, not bare, quantities.
\end{references}
\end{document}